# Matching domain wall configuration and spin-orbit torques for very efficient domain-wall motion

A.V. Khvalkovskiy[1], V. Cros[2], D. Apalkov[1], V. Nikitin[1], M. Krounbi[1], K.A. Zvezdin[3], A. Anane[2], J. Grollier[2], A. Fert[2]

[1] Grandis, Inc., San Jose, California, U.S.A

[2] Unité Mixte de Physique CNRS/Thales and Université Paris Sud, Palaiseau, France

[3] A.M. Prokhorov General Physics Institute, Russian Academy of Sciences, Moscow, Russia

Abstract :

In our numerical study, we identify the best conditions for efficient domain wall motion by spin-orbit torques originating from the Spin Hall effect or Rashba effect. We demonstrate that the effect depends critically on the domain wall configuration, the current injection scheme and the symmetry of the spin-orbit torque. The best identified configuration corresponds to a Néel wall driven by spin Hall Effect in a narrow strip with perpendicular magnetic anisotropy. In this case, the domain wall velocity can be a factor of 10 larger than that for the conventional current-in-plane spin-transfer torque.

The fundamental concepts of spintronics are based on the generation, manipulation and detection of spin polarized currents. It has led in the last two decades to the development of a new generation of magnetic sensors, and notably read heads of hard disks and of non volatile magnetic memories (MRAM) that are expected to supplant semiconductor based memory devices. In classical spintronics, one generally creates a spin current by passing a charge current through a thin ferromagnetic layer, whose magnetization direction can be controlled by an external applied field or more efficiently by spin transfer torques. Recently an alternative way has emerged to control the magnetization configuration of a ferromagnetic layer that is based on current-induced spin-orbit (SO) torques, namely the Spin Hall Effects [1] and the Rashba effect [2]. Such SO torques are expected for magnetic stripes adjacent to a nonmagnetic conductive layer with strong spin-orbit interactions (*SO layer*) [3]. The exploitation of these SO effects opens up alluring possibilities for the development of a novel generation of highly reliable and low energy spintronic devices. In particular, several experiments [4-8] have already demonstrated the great interest of the SHE and/or Rashba effects in metallic nanostructures to switch the magnetization of a single nanomagnet.

In this work, we discuss the relevance of these SO effects for obtaining the most efficient manipulation of Domain Walls (DW) in magnetic nanoribbons. We use micromagnetic simulations to explore the impact of SO torques on the DW motion for both in-plane and perpendicularly magnetized magnetic layers. We consider different DW types i.e. Bloch wall (BW), Néel wall (NW) in perpendicularly magnetized materials or Head-to-Head DW (HTH) and relative orientations of the electron flow and the orientation of the magnetic stripe. We find that some of these geometrical configurations are particularly promising for fast DW motion induced by spin-orbit torques with small current densities.

We first introduce the two types of SO torques under investigation here. As for the Spin Hall Effect, an interfacial spin accumulation is created through the imbalance of the deflections of spin up and spin down electrons away from the electric field direction in a non magnetic layer [1]. The resulting transverse pure spin current can be injected into an adjacent magnetic layer in order to exert a spin transfer torque (STT) and manipulate the magnetization. The Spin Hall angle $\Phi_H$ is the parameter characterizing the maximum yield for the conversion of a longitudinal charge current density into a transverse spin current density:

$$j_s = \Phi_H j_c/e \qquad (1)$$

where $j_c$ is the charge current and *e* the electron charge. After the initial predictions by Dyakonov and Perel, the first experimental observations in pure metals [9] or semiconductors [10] have confirmed either by optical means or magnetotransport measurements, the existence of spin currents associated to SHE. However the deduced spin Hall angles had relatively modest values. Interestingly, very recently, large values of the SH angle have been reported in metals and alloys, up to $\Phi_H$ = 0.11- 0.15 in Ta [6], $\Phi_H$ = 0.24 in Cu doped with Bi impurities [12], and $\Phi_H$ = 0.33 in W [13], which shifts the interest in SHE from an exotic transport phenomena to an effective driving force for magnetization manipulation.

The second spin orbit related phenomena is the Rasbha effect [2,3], which is induced by the electric field gradient due to a symmetry breaking at a surface or interface, typically a system composed of SO-Metal/FerroM/Oxide. It results in a coupling by exchange between spin and momentum in surface or interface states expressed by the Rashba Hamiltonian:

$$\hat{H}_R = \alpha_R \vec{s}.(\vec{k} \times \hat{e}_z) \qquad (2)$$

where $\vec{s}$ is the spin, $\alpha_R$ is the Rashba parameter, $\hbar\vec{k}$ the electron's momentum and $\hat{e}_z$ a unit vector perpendicular to the surface or interface. Experimentally, Rashba spin-orbit splittings have been measured by ARPES only for a few metallic or semi-conductor interfaces states and the extracted Rashba parameter remain small with $\alpha_R$ values in the range of $10^{-11}$eV.m. Remarkably, a very large value has been observed at Bi/Ag interfaces, $\alpha_R = 3\times10^{-9}$eVm, making Bi based systems particularly attractive for spin-orbit driven control of the magnetization [14]. In the presence of the Rashba interaction, the charge current flowing in a magnetic thin layer in a direction parallel to the interface generates a spin accumulation that can interact by exchange with the magnetic moment of the adjacent magnetic material. Recent experiments by I.M. Miron et al suggest that equivalent Rashba fields of $\approx$ 3 to 10 kOe can be created by this interaction [4-5]. In addition to this effective field, the Rashba-induced spin accumulation can give rise to an additional term of spin transfer torque, called later on "Indirect Rashba Effect" (IRE). This new term has been predicted to arise either from the spin diffusion inside the magnetic layer or from a spin current associated to Rashba interaction at the interfaces with the spin-orbit metal [15-16]. The symmetry and the influence of these new contributions among the spin-orbit transfer torques will be discussed later.

The main objective of this work is to investigate the effect of SHE and Rashba torques acting on three different types of DW that are head to head (HTH) DW in magnetic nanostrips with in-plane magnetization, Bloch wall and Néel DWs in out-of-plane magnetized nanostrips (see Fig.1 a). In our simulations, we have used structures that are bilayers consisting of: (i) a ferromagnetic nanostrip called (F), (ii) a nanostrip made of a material with large spin-orbit (SO) coupling. In order to address all possible symmetries, we have considered two different sample geometries. First, the parallel geometry (see Fig.1b) corresponds to the most common geometry studied in the experiments for which both the SO layer and the magnetic material are stacked in the same nanostrip. The second case corresponds to perpendicular nanostrips (see Fig 1c) for which the different SO torques will be active only at the intersection of the two layers. To our knowledge, such perpendicular geometry has never been explored.

From now, we aim at analyzing the influence of the torques generated either by spin currents resulting from a combination of SHE and IRE or by the coupling of the magnetization of the F layer with the Rashba field. It is also important to precise that the direction of the spin accumulation in the SO layer depends on the direction of the charge current that flows along the magnetic wire axis in the parallel geometry (Fig.1b) and perpendicular to the magnetic wire in perpendicular geometry (Fig.1c). Note that we use here the same approach as the one we proposed to highlight the crucial advantage of perpendicular spin current injection to generate large DW velocities considering classical spin transfer torques [17]. Finally, the two possible symmetries for the spin-orbit torques per unit moment that are acting on the local magnetization correspond to the following expressions for the time derivative of the unit vector $\hat{m}$ along the magnetization axis:

$$\vec{\tau}_1 = -\gamma \tau_1 (\hat{m} \times \hat{\sigma} \times \hat{m}) \quad (3)$$

$$\vec{\tau}_2 = -\gamma \tau_2 (\hat{m} \times \hat{\sigma}) \quad (4)$$

$\hat{\sigma} = \hat{j} \times \hat{z}$, $\hat{j}$ and $\hat{z}$ are unit vectors in the current and out of plane directions and $\gamma$ is the gyromagnetic ratio. The torque $\vec{\tau}_1$ in Eq. 3 presents a symmetry similar to the classical Slonczewski torque or in-plane torque generated by the injected spin current density $j$. As far as the role of Spin Hall Effect is considered, we suppose for simplicity that this spin current density has its optimal value, i.e. $\Phi_H j_c / e$ as expressed in Eq.1. The resulting torque magnitude $\tau_1$ (in unit moment) can be written as:

$$\tau_1^{SHE} = \frac{\hbar \Phi_H j_c}{2 e M_S t_F} \quad (5)$$

where $t_F$ is the thickness of the ferromagnetic layer (F) and $M_S$ its saturation magnetization. The second torque $\vec{\tau}_2$ in Eq.4 has the same symmetry as the classical Field Like Torque (FLT) or also called perpendicular torque. It is now generally admitted that this second torque remains very weak in metallic systems [17], whereas very large perpendicular torques are predicted in magnetic tunnel junctions [18] and confirmed experimentally with large DW velocities induced by spin torques with perpendicular injection [19]. Consequently, as the system we study here is composed of metallic layers, one gets:

$$\tau_2^{SHE} \ll \tau_1^{SHE} \quad (6)$$

If we now consider that the torque is due to the Rashba effect (RE), one generally considers only the effect due to the Rashba effective field $\vec{H}_{Rashba}$ due to the direct exchange like interaction between the Rashba-induced spin accumulation and the magnetization [3]. The resulting torque has the symmetry of Eq. (4), and its magnitude $\tau_2$ is given by approximately [20]:

$$\tau_2^{Rashba} = \frac{\alpha_R j_c P}{\mu_B M_S} \quad (7)$$

Where $\mu_B$ is the Bohr magneton and P is the polarization of the carriers in the ferromagnetic layer; the Rashba parameter $\alpha_R$ is averaged over the magnetic film thickness. As already mentioned, an "indirect Rashba effect" has been predicted and results in a torque of the SLT symmetry acting on the ferromagnetic layer.

In order to predict their influence, we have performed micromagnetic simulations of the DW motions induced by both torque symmetries. From now on, we consider that the SLT torque comes primarily from the SHE and/or IRE while theFLT comes primarily from the direct Rashba effect. The effective fields associated with these torques can be defined conventionally:

$$\vec{H}_{SLT} = \tau_1 (\hat{m} \times \hat{\sigma})$$

$$(8)$$

$$\vec{H}_{FLT} = \tau_2 \hat{\sigma} \quad (9)$$

Where $\tau_1$ and $\tau_2$ are can be related respectively to the expressions Eq. (5) and Eq. (7) for the torques for the SHE and direct RE.

These two spin torques symmetries have been predicted to strongly influence the DW dynamics [21]. Here we propose to investigate the conditions in terms of DW types and injection scheme (parallel geometry in Fig. 1b and perpendicular geometry 1c) under which one of theses spin-orbit torques is able to generate a very fast DW motion. The simulations are performed for a long and thin (4 micrometer long and 3 nm thick) magnetic strips with different domain wall (DW) configurations: (i) in plane magnetized NiFe layer with Head-to-Head (HTH) transverse walls, (ii) perpendicularly magnetized Co/Ni multilayer with either Bloch or Néel DWs. For NiFe, we use the following magnetic parameters: $M_s$ = 800 emu/cc, A = 1.3e-6 erg/cm2, $\alpha$ = 0.01. The width of the in-plane magnetized strip is chosen to be 100 nm. For simulating the case of perpendicularly magnetized layers, we use magnetic parameters corresponding to Co/Ni i.e. $M_s$ = 650 emu/cc, A = 2e-6 erg/cm2, $K_{perp}$ = 3.3e6 erg/cc, $\alpha$ = 0.02. In order to get the two DW structures, Bloch wall and Néel wall, we perform the simulations for two different widths: w= 200 nm for Bloch DW and w = 50 nm for Néel DW. For the parallel configuration (see Fig. 1b), the SO and the magnetic layers are adjacent on the whole length of the strip. For the perpendicular configuration (see Fig. 1c), we chose a strip width of 200 nm for the NM-SO layer that is perpendicular to the magnetic strip. For the micromagnetic simulations, we start with an equilibrium state with a DW initially located in the middle of the magnetic strip for the parallel configuration and centred at the axis of the non magnetic SO strip for the perpendicular one. At time t = 0, a dc current is switched on, that in turn induces the SO torques acting on the DW. The magnitude of the torque and the field of the SHE is calculated using $\Phi_H$ = 0.2. The field of the RE (see Eq. (9)) is taken to match that of the SHE Eq. (8) (for $\hat{m} \perp \hat{\sigma}$). Therefore the magnitude of both the effective fields is about 3 Oe at $j_c$ = $10^6$ A/cm$^2$ for both NiFe and Co/Ni The micromagnetic simulations are performed by numerical integration of the LLG equation using our SpinPM micromagnetic package based on the fourth order Runge-Kutta method with an adaptive time-step control for the time integration. The Oersted field contribution is not considered in the numerical simulations. Generally, it has the same symmetry as the Rashba field, with the only difference that the Rashba field can have different signs for different interfaces. However, the Oersted field and the Spin Hall/Rashba effects scale differently with the system size: at a given current density, the Oersted field contribution becomes negligible for small sizes, while the magnitude of the SHE and RE does not change. .

We first investigate how the different SO effects are able to displace a DW in the parallel geometry, that is the one commonly used in all the existing experimental studies [4-8]. In Fig 2, we display the results for the different DW types and for a current density J = 1 MA/cm2. The most remarkable result is that the only case in which a DW can be brought into a steady and fast motion (without any additional applied field) is for a Néel DW under the action of the torque due to SHE (red curve and inset in Fig. 2). The torque associated to RE only causes a small and reversible DW shift (blue curve in Fig. 2). For the two other DW types, none of the SO torques is able to induced substantial DW displacements (green curve in Fig 2). This result can be simply explained if one considers the orientation of the effective field acting on the DW, which must be aligned with the magnetization of one of the adjacent domains to promote a steady motion [17]. For the parallel geometry, this prerequisite is fulfilled only for the Néel DW in perpendicularly magnetized strip excited by the SHE: the effective field $H_{SLT}$ is perpendicular to the plane at the center of the DW (see Eq. 4). In contrast, in case of Bloch and HTH DWs, the directions of the SO spin accumulation and the magnetization direction at the DW center are parallel to each other along Oy and thus $H_{SLT}$ vanishes. As for the Rashba effect, the direction of the effective field, $H_{Rashba}$, according to Eq. (5), is given by the direction of the SO polarization $\hat{\sigma}$, which is along Oy, and therefore it is not along the domain direction for all three DW types.

Our results of steady and fast motion with Néel walls might appear rather surprising as Bloch DWs are expected in the existing experiments on the rather large Co/Pt or Co/Ni nanostrips in which large velocities have been obtained and ascribed to SHE or Rashba effects [4-7]. Indeed, Haazen et al have recently observed that a DW motion is induced by SHE in asymmetric Pt/Co/Pt strip only when they apply a large field along Ox needed to favor Néel DWs [8]. It has been also shown that the Dzyaloshinski-Moriya (DM) interaction can be large at the interfaces with the materials having large SO interaction. The DM interaction can result in stabilization of the Néel DW structure over the Bloch DW structure even for rather wide stripes [22]. Taking into account such DM coupling, A. Thiaville et al recently demonstrate that it influences not only the static DW structure by favoring Néel DW but also the DW dynamics as it makes more difficult the transition to a Bloch DW structure, which can result in shifting the Walker breakdown and therefore allowing large DW velocities[23]. Therefore very large DW velocities observed recently in wide nanostrips made of Co/Pt and Co/Ni [4-8] can be accounted for by the SHE or the indirect RE.

We now focus on the second geometry with the perpendicular configuration of the current (depicted in Fig. 1(c)). In Fig. 3a, we demonstrate that a DW steady motion is obtained for two combinations of DW type/SO effect (i) a HTH DW excited by the direct Rashba effect (ii) a Bloch DW excited by the SHE. We note that both the amplitude and the evolution of DW displacement with current depend on the DW type, with a largest value obtained for HTH DW driven by RE. These results can be explained, as in the case of parallel geometry, by looking at the directions of the effective fields associated with each SO torque type acting on the DW and compare them to magnetization directions in the adjacent domains. Since the polarization vector $\hat{\sigma}$ for the perpendicular geometry is along the $O_x$ direction (see Fig. 1(c)), the Rashba field is also directed along $O_x$. As a consequence, $H_{Rashba}$ is parallel to one of the magnetic domains for the case of in-plane magnetized strip, explaining why Rashba-induced torque is able to put a HTH DW into motion in this perpendicular configuration (red curve in Fig. 3 (a)). To our knowledge, this case has never been studied. As for the two other DW types i.e. Bloch and Néel DW that are of some interest for memory applications in, as the Rashba field $H_{Rashba}$ is always perpendicular to the domains, we conclude that both DWs are only shifted on a very small distance of the order of a few nm. If we now consider the action of SHE, its effective field $H_{SLT}$ is along $O_z$ for the in-plane magnetized strip and vanishes for the perpendicular geometry with the Néel DW, and thus no DW motion is expected for these cases. Finally, we find that the SHE is able to move a Bloch DW in this geometry again because $H_{SLT}$ is perpendicular to the plane.

In the perpendicular geometry, the SO-induced torque acts on the DW only within the intersection of the magnetic strip and the SO current line. Therefore, there can be no steady-state DW motion in this configuration. Instead, three regimes of the DW motion are observed; see Fig. 3(b). First, when the DW is within the intersection, it is accelerated by the SO-field. DW gets its maximum velocity just before leaving the intersection region. The second regime is the DW inertial motion after it leaves the intersection; now the DW is moving with the velocity decreasing linearly due to the damping. Eventually, the DW stops or it can start to move with a small velocity if it is attracted or repelled by the edges of the stripe. The maximum DW velocity as well as the DW total shift during this damped motion regime scale with the current density, as shown in Fig. 3(b). When the damping is sufficiently small and there are no pinning sites in the wire, we find that the DW can reach few hundred m/s in velocities and thus can travel on a distance which is few times larger than the length of the intersection region. Practical applications can be designed so that the magnetic strip has two or several intersections with the SO current lines, and a DW can be moved from one SO line to another by a current pulse flowing in the first SO line. In Table 1, we summarize all our results of SO driven DW motion in the different geometries.

Now we compare the efficiency of the DW motion by the SO-induced torques to the conventional spin transfer torques for the in-plane currents. The DW velocity of a Néel DW driven by SHE reaches about ~ 20 m/s at current density of $j = 0.2$ MA/cm² and increases linearly with $j$, (see inset of Fig. 2). The maximum velocity we find for the Bloch DW in the perpendicular configuration for the SHE is about 7 m/s for the same current. These values are close to the prediction of the 1D model for a uniform external field $V_{DW} = \dfrac{\gamma H_{eff} \Delta}{\alpha}$, where $\Delta$ is the DW width, $\gamma$ the gyromagnetic ratio and $\alpha$ the damping. In our case, the effective field $H_{eff}$ to $H_{SHE}$. In the conventional CIP case, one can use the same equation, where now $H_{eff}$ is equal to effective field associated to the non adiabatic torque [24] : $H_{IP} = \dfrac{\hbar J_{IP} \beta P}{2e\Delta}$, where $\beta$ is the non adiabatic parameter. Comparing this expression to the Eq. (5-6), which give the amplitude of the SO fields for either cases i.e. $H_{SO} = \dfrac{\hbar J_{IP} \Phi_{SO}}{2ed}$, we easily find that the ratio of the SO field and CIP field is equal to the ratio of the DW velocities: $\dfrac{v_{SO}}{v_{CIP}} = \dfrac{H_{SO}}{H_{CIP}} = \dfrac{\Phi_{SO}}{\beta P} \dfrac{\Delta}{d}$. Thus we conclude that SO driven DW motion can be faster than in CIP case by at least a factor of 10 as the non adiabatic term $\beta$ is always about a few percent [25] and the ratio DW width over film thickness $\Delta/d$ never lower than unity.

In conclusion, we have performed a comprehensive numerical study to identify the best conditions of domain wall type, current direction and spin-orbit induced transfer torque, SHE or Rashba promoting a fast current-induced DW motion. As this can be equally explained by symmetry arguments, we find that DWs can be moved using SO effects in three cases : (1) in a perpendicularly magnetized strip with current parallel to the strip and SHE torque (2) Bloch wall (out of-plane magnetization) by SHE and Head-to-Head DW (in-plane) by Rashba effect with a perpendicular current. Moreover, we find an improved efficiency of SO torques comparing to the conventional CIP DW motion, making the SO effects very promising to be used in the next generation of DW based spintronic devices.

Acknowledgements:

This work is partly supported by the ANR agency (grant "ESPERADO" 11-BS10-008 and SpinHall), the RFBR (grant 11-02-91067) and CNRS PICS Russie (Grant No. 57432011).

Figure captions :

Fig. 1: (a) Schematic view of the three different DW types that have been considered i.e. Head to Head DW in an in-plane magnetized strip, Bloch or Néel DW out-of-plane magnetized strip (b) Parallel geometry in which the ferromagnetic layer (F) is parallel to the spin-orbit (SO) layer on the whole length of the strip (c) Perpendicular geometry for which the SO strip is perpendicular to the F strip.

Fig. 2: DW shift as a function of time under different possible SO torques in the parallel geometry. A steady motion is found for the case of a Néel wall excited by SHE (red curve) whereas for Néel wall with RE, the DW is only shifted (blue curve). (Inset) Néel DW velocity induced by SHE versus current density in parallel geometry

Fig. 3: DW velocity as a function of DW position inside the F strip. Zero corresponds to the center of the intersection with the SO strip that is 200nm wide. A steady motion is found for Bloch DW excited by SHE and for HTH DW excited by RE. Two regimes are identified i.e. (I) the DW is inside the intersection area and a steady motion is obtained (II) the DW is outside the intersection and an inertial motion with decreasing velocity is found

Table 1 : Summary of the numerical results for the different DW types and injection geometry

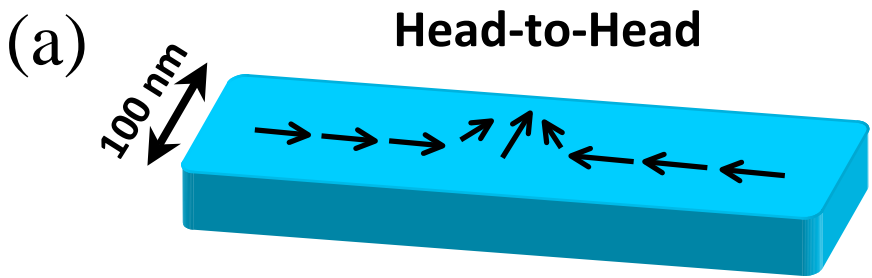
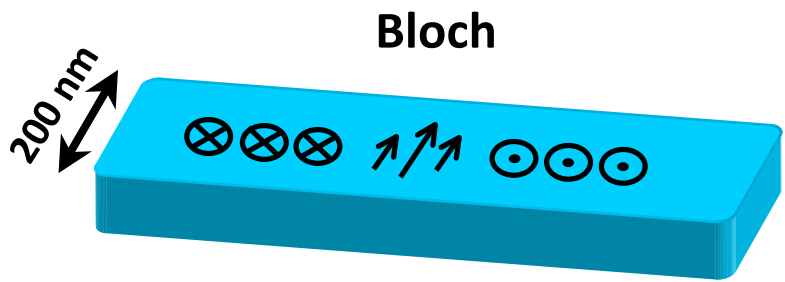
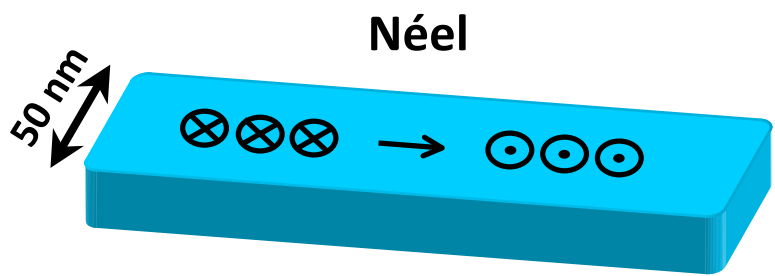
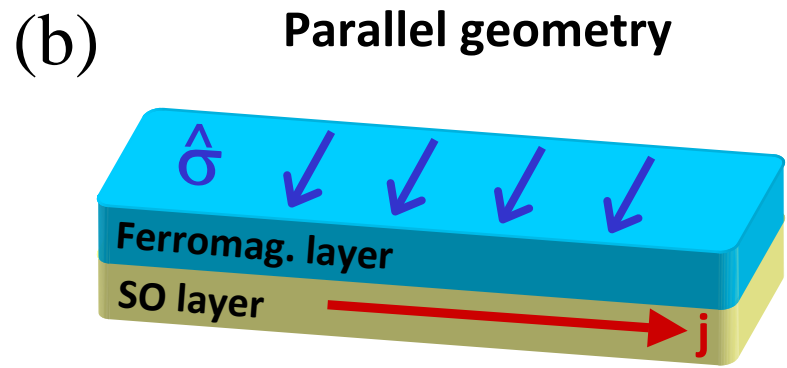
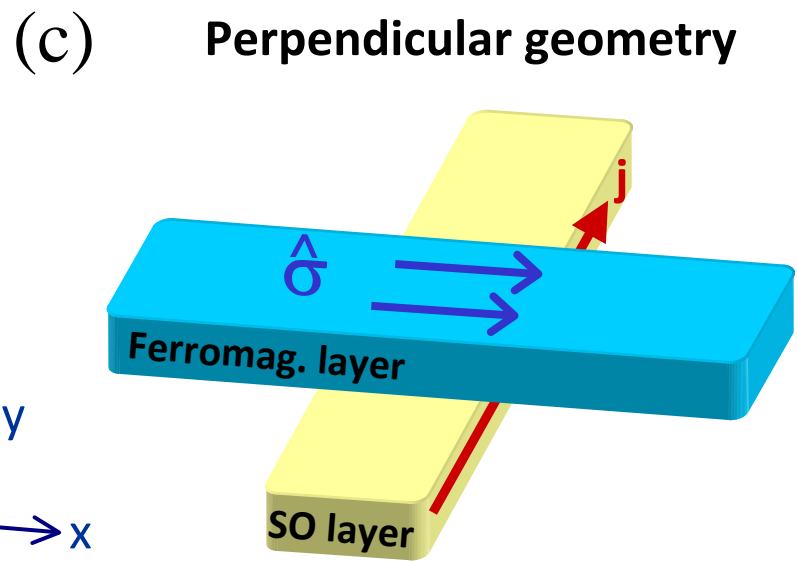

Fig. 1

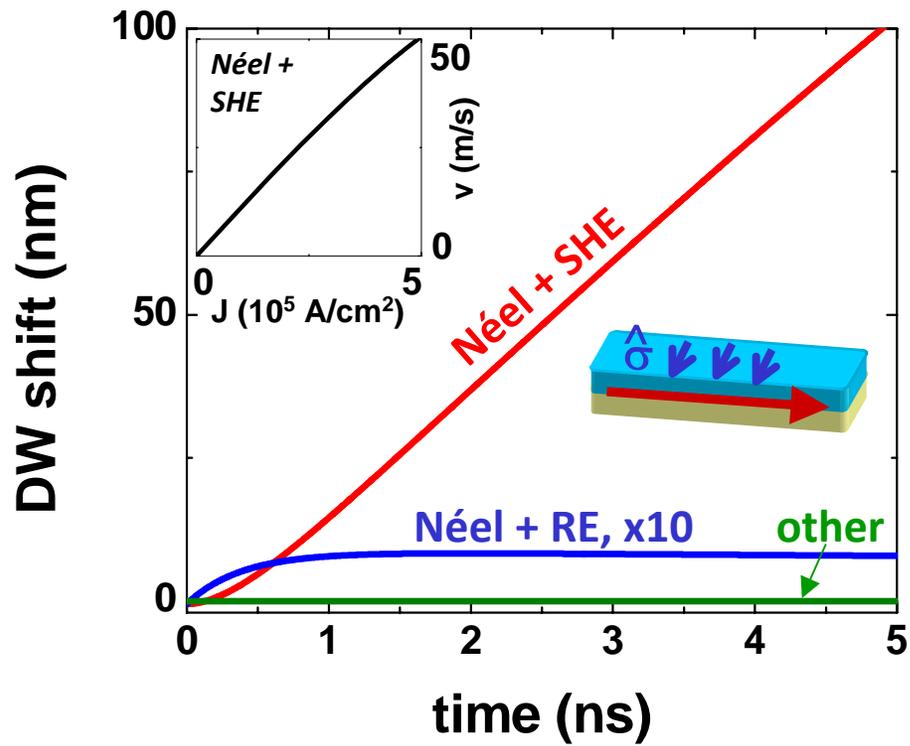

Fig. 2

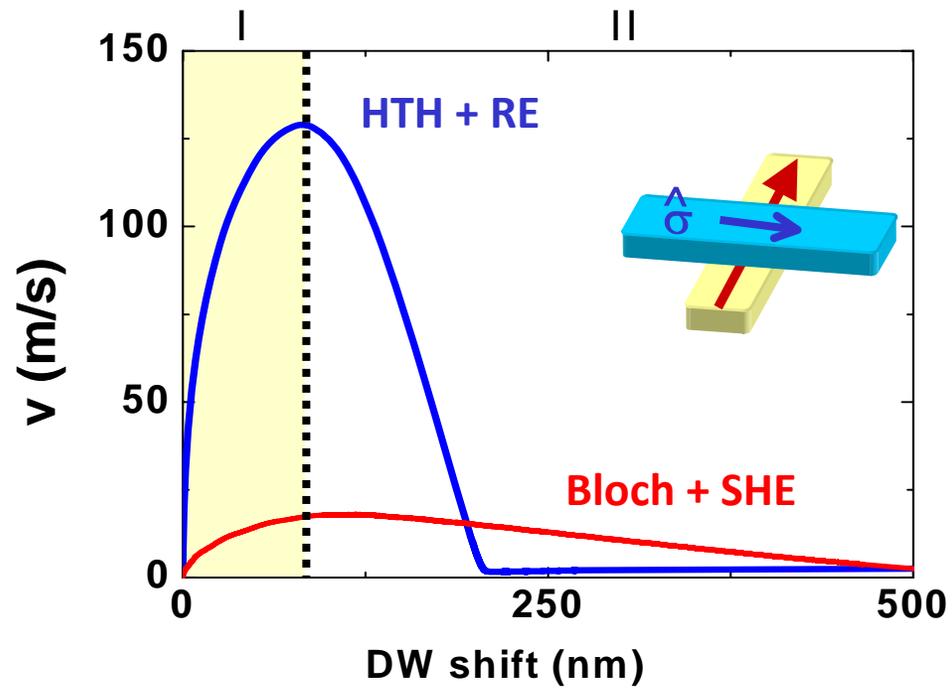

Fig. 3

| Injection geometry | Torque origin | DW type | Result |
|---|---|---|---|
| parallel 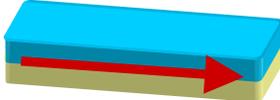 | SHE and/or IRE | Bloch | no motion |
| | | Néel | **steady motion** |
| | | HTH | no motion |
| | Rashba | Bloch | no motion |
| | | Néel | shift |
| | | HTH | no motion |
| perpendicular 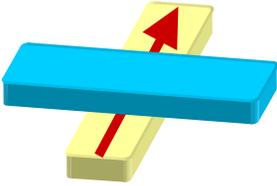 | SHE and/or IRE | Bloch | **steady motion** |
| | | Néel | no motion |
| | | HTH | no motion |
| | Rashba | Bloch | shift |
| | | Néel | no motion |
| | | HTH | **steady motion** |

Table. 1